\documentclass{aa}  
\usepackage{xcolor}

\usepackage{longtable, booktabs}
\usepackage[breaklinks, colorlinks, citecolor=blue]{hyperref}

\usepackage{graphicx}
\usepackage{txfonts}
\usepackage{bm}
\usepackage{comment}
\usepackage{multirow}
\def\drm{\mathrm{d}}
\def\los{{\bf \hat n}}
\def\kvec{{\bf  k}}
\def\kvecunit{{\bf \hat k}}
\def\Pee{P_{ee}}
\def\qvec{{\bf \tilde {q}}}

\def\exp{{\rm e}}

\begin{document}

   \title{A consistent, physical, and analytical model for \\CMB observables of reionisation}

   \subtitle{I. Derivation, constraining power, and detectability of the auto-spectra}

   \author{Adélie Gorce\thanks{adelie.gorce@universite-paris-saclay.fr}
          }

   \institute{
Université Paris-Saclay, CNRS, Institut d’Astrophysique Spatiale, 91405, Orsay, France
\label{inst1}
             }

   \date{Received ...; accepted ...}

 \abstract{The Epoch of Reionisation imprints its history and morphology on the Cosmic Microwave Background temperature and polarisation anisotropies through two effects: The kinetic Sunyaev Zel'dovich (kSZ) effect and Thomson scattering. To perform joint analyses of these observables and put tight constraints on the reionisation history and morphology, a common reionisation model is necessary. We present a quick, physically-accurate analytical approach to derive consistently the angular power spectra of the three main resulting imprints that are the spatial fluctuations of the Thomson optical depth, the patchy kSZ effect, and the scattering and screening $B$-modes. The approach differs from existing analytical models, as it is calibrated on high-resolution hydrodynamical simulations and depends on physical parameters of reionisation. Our predicted power spectra are consistent with the ones derived from cosmological simulations, whilst including more physical information -- such as the typical size of ionised bubbles and the duration of reionisation. We provide an updated detectability assessment and show that both $C_\ell^{\tau\tau}$ and $C_\ell^{BB}$ could be measured by experiments similar to CMB Stage 4 and CMB-HD, respectively.
 We show the complementarity of the three observables in constraining reionisation history and morphology, and illustrate the potential of their joint analysis to get a global picture of reionisation. This analytical model will be a powerful tool in the analysis of upcoming CMB data, either alone or in combination with independent datasets such as measurements of the high-redshift 21\,cm signal.
 The code to derive these spectra is publicly available online as the \texttt{preion} Python package.}


   \maketitle
%

\section{Introduction}

Reionisation is a patchy process, that is, different parts of the sky got reionised at different points in time \citep{Aghanim96}. According to recent simulations, the favoured scenario is one where the densest regions of the IGM, which host the ionising sources, get reionised first. Starting from there, the ionised bubbles surrounding sources slowly grow and percolate until the entirety of the IGM is ionised.

The interaction of the free electrons from reionisation and the CMB photons produces specific secondary anisotropies. First, the local CMB quadrupole Thomson scatter off the free electrons, generating polarisation. This effect increases the amplitude of the polarisation anisotropies on the scales corresponding to the horizon then ($\ell < 10$) and is seen as a `reionisation bump', the amplitude of which on the $EE$ power spectrum scales as $\tau^2$, where $\tau$ is the Thomson optical depth. Because the distribution of free electrons is inhomogeneous, this interaction also produces $B$-modes with a characteristic imprint on the polarisation power spectrum. 
Second, incoming radiation is scattered in and out of the line-of-sight in a process called `patchy screening'. This leads to dampened temperature fluctuations on small scales as $\mathrm{e}^{-2\tau}$ ($\ell > 200$).
Finally, CMB photons can get Doppler boosted by high energy electrons. These electrons can be either hotter than the CMB, or have a proper velocity with respect to the CMB rest-frame. The resulting effects are called, respectively, thermal and kinetic Sunyaev Zel’dovich, and are most significant on small scales \citep[$\ell > 1000$,][]{ZeldovichSunyaev_1969, SunyaevZeldovich_1980}. The contribution of reionisation to the tSZ effect is negligible compared to the contribution of galaxy clusters, much denser and warmer than ionised bubbles \citep[see][for a recent discussion of this contribution]{IlievHosein_2025}. However, the `patchy' part of the kSZ signal, coming from the Epoch of Reionisation, has sufficient amplitude and shape to be separated from the low-$z$ component and has already been used to put constraints on the reionisation history \citep[e.g.,][]{planck_2016_reio,zahn_2012_spt,ReichardtPatil_2021,GorceDouspis_2022, BeringueSurrao_2025, ChaubalHuang_2026}.

In order to use its imprints on the CMB power spectra to constrain reionisation, a model is required. However, modelling cosmic reionisation is challenging, because of the wide range of time and physical scales involved, and several approaches have been considered in the literature. High-resolution numerical simulations provide the most physically detailed models, as they solve the three-dimensional radiative transfer equation to track photons from their source galaxy into the IGM \citep{Aubert2015_EMMA, GaraldiKannan_2024, GiriBianco_2024}. However, such simulations are extremely costly to run, which limits the comoving volumes they can cover to order $(100~\mathrm{Mpc})^3$, as well as the number of models which can be explored (although see \citet{MeriotSemelin_2024} for an intermediate approach allowing the authors to produce a 10~000 simulation database). Semi-numerical simulations partially solve these issues by modelling the non-linear matter distribution on large scales whilst relying on calibrated prescriptions to approximating small-scale astrophysical processes \citep{ChoudhuryParanjape_2021, TracChen_2022, DaviesMesinger_2025}. Their use in an inference framework is, however, again limited to moderately-sized simulation cubes (\citealt{QinPoulin_2020,ChoudhuryMukherjee_2021,NikolicMesinger_2023}, although see \citealt{ChenTrac_2023}), insufficient to track the large scales involved in both the kSZ and Thomson scattering of CMB photons during reionisation \citep{Alvarez_2016}. Finally, the most computationally-efficient approach, which can cover extremely large scales, is the analytical approach. This is the approach we focus on in this work.

Starting from a parameterisation of the free electron overdensity power spectrum across scales and times, introduced in \citet{GorceIlic_2020}, we propose an analytical framework to derive consistently all the imprints of reionisation on the CMB temperature and polarisation power spectra, from large ($\ell<10$) to small ($\ell>1000$) scales, in Sec.~\ref{sec:methods}. This framework will be made publicly available as the \texttt{preion} Python package upon publication of this manuscript. In Sec.~\ref{sec:results_ps}, we investigate the dependency of these imprints on the model parameters describing both the global history of reionisation ($z_\mathrm{re}$, d$z$) and its morphology (log$\alpha_0$, $\kappa$), demonstrating their potential to measure such parameters. In Sec.~\ref{sec:potential}, we assess the detectability of each imprint, with current and future, space- and ground-based CMB experiments. We conclude and discuss our findings in Sec.~\ref{sec:conclusions}.

\section{Methods: Analytical power spectrum}\label{sec:methods}

In this section, we describe how to derive the angular power spectra of the patchy kinetic SZ effect, of the fluctuations of the Thomson optical depth, and of the EoR-induced $B$-mode anisotropies, given an analytical model of reionisation. These derivations are implemented in a publicly available Python package called \texttt{preion}\footnote{At \url{https://github.com/adeliegorce/preion}.}.

\subsection{Reionisation histories}\label{subsec:2_reion}

The first ingredient of our analytical model is  a parameterisation of the volume-averaged reionisation of \ion{H}{I} and \ion{He}{I} \citep{Douspis2015,planck_2016_reio,GorceDouspis_2018}:
\begin{equation}
\label{eq:xe_param}
x_e(z)  = 
\left\{ 
	\begin{array}{ll}
		f_H & \mathrm{for} \: z< z_\mathrm{end},\\
		f_H\, \left(\frac{z_\mathrm{early}-z}{z_\mathrm{early}-z_\mathrm{end}}\right)^\alpha & \mathrm{for} \: z>z_\mathrm{end},
	\end{array}
\right.
\end{equation}
where $z_\mathrm{early}=20$ corresponds to the redshift around which the first emitting sources form and at which $x_e(z)$ is matched to a fixed, cosmology-independent residual ionised fraction of $10^{-4}$, and $f_H$ is the fractional quantity of electrons per Hydrogen atom. The reionisation midpoint $z_\mathrm{re}$ corresponds to the time when $50\%$ of the Hydrogen in the Universe is ionised, whilst its endpoint $z_\mathrm{end}$ is reached when $100\%$ of it is ionised. This parameterisation reproduces the behaviour observed in various simulations of reionisation \citep{Aubert2015_EMMA, Seiler2019_rsage, 21cmfast_v3, MeriotSemelin_2024}, that is a slow start, as the first ionising sources light up, before a power-law acceleration, until full Hydrogen ionisation is reached\footnote{For compatibility, the \texttt{preion} package also allows the exponential parameterisation introduced in latest version of the Boltzmann code \texttt{CAMB} \cite{camb1, camb2}.}.
The parameters of the model are $z_\mathrm{re}$ and $z_\mathrm{end}$, Varying $z_\mathrm{re}$ is equivalent to varying $\alpha$ and to changing the slope of the redshift-evolution. We consider the first reionisation of Helium to be simultaneous with the Hydrogen's ($f_H\simeq 1.08$).

We include the second reionisation of Helium as a redshift-symmetric `instantaneous' process, described by an hyperbolic tangent \citep{camb1,camb2}, such that:
\begin{equation}
\label{eq:xe_tanh}
x_\ion{He}{III}(z) = \frac{1}{2}\, \left[ 1 + \text{tanh}\left( \frac{y-y_\mathrm{re}}{\delta y}\right) \right], 
\end{equation}
where $y\, (z)\equiv(1+z)^{\frac{3}{2}}$, $y_\mathrm{re}=y\, (z=z_\mathrm{re, HeIII})$ for $z_\mathrm{re, HeIII}=5.0$ the midpoint of $\ion{He}{III}$ reionisation, and $\delta y \equiv \frac{3}{2}\, (1+z)^{\frac{1}{2}}\, \delta z$ for $\delta z = 0.5$, roughly corresponding to the redshift range which sees the $\ion{He}{III}$ fraction increase from $25\%$ to $75\%$ \citep{kuhlen_2012}. 

Implemented in the Boltzmann code \texttt{CAMB}, this parameterisation is used to derive the large-scale reionisation bump in the CMB temperature and polarisation power spectra.

\subsection{The power spectrum of electron density fluctuations}\label{subsec:2_Pee}

Numerous previous works have used effective, analytical derivations of the power spectra of EoR imprints \citep[e.g.,][]{DvorkinSmith_2009}. The most commonly used model was introduced in \cite{FurlanettoZaldarriaga_2004} and consists in parameterising the ionised bubble radii as a log-normal distribution whose parameters are constant with redshift. This parameterisation is then used to derive analytic expressions of $P_{ee}$, assuming it is a biased tracer of the dark matter power spectrum. In analogy to the halo model, the electron power spectrum is divided into a one-bubble and a two-bubble term \citep{WangHu_2006}. The resulting $P_{ee}$ can, then, be used in Eqs.~\eqref{eq:cl_tau_Pee}, \eqref{eq:def_C_ell_pkSZ}, and \eqref{eq:cl_bb_Pee}. However, the reionisation field is more complex than allowed by this model, with bubbles merging and overlapping as soon as $10\%$ of the IGM is ionised to create a network of $\ion{H}{II}$ regions which eventually grows to cover the entire IGM. Such models also ignore the possibility for partially ionised regions to exist, which could affect the observables \citep{SobacchiMesinger_2014}, and become less accurate in the non-linear regime (especially on the small scales probed by, among others, the kSZ signal). 

In this work, we use the physically-motivated model introduced in \citet{GorceIlic_2020} to describe the time- and scale-evolution of the power spectrum of free electron density fluctuations. The model was calibrated on high-resolution fully hydrodynamical simulations of reionisation \citep{Aubert2015_EMMA} and tested against semi-numerical simulations \citep{21cmfast_v3,Seiler2019_rsage}. The parameterisation splits the redshift-evolution of $P_{ee}(k,z)$ into two components. The high-redshift part is described by a power-law and includes the reionisation morphological information, whilst the low-redshift component is described by a biased matter power spectrum and dominates post-reionisation.

The overall model writes:
\begin{equation}
    \label{eq:Pee_model}
\begin{aligned}
P_{e e}(k, z)=&\left[f_{\mathrm{H}}-\bar{x}_e(z)\right]  \times \frac{\alpha_0 x_e(z)^{-1 / 5}}{1+[k / \kappa]^3 x_e(z)} \\
& +x_e(z)\, b_{\delta b}(k, z)^2\, P_{\delta \delta}(k, z),
\end{aligned}    
\end{equation}
where $\kappa$ is the electron drop-off frequency and $\alpha_0$ is the (constant) large-scale $P_{ee}$ amplitude. We have shown in \cite{GorceIlic_2020} that both $\alpha_0$ and $\kappa$ can be related to the morphology of reionisation: Statistically larger ionised bubbles during reionisation will be related to smaller values of $\kappa$, whilst numerous, small ionised bubbles will lead to a low-variance electron density field, and, in turn, to low values of $\alpha_0$.
The low-redshift component of equation~\eqref{eq:Pee_model} corresponds to a biased non-linear matter power spectrum. The redshift-independent baryon-dark matter bias $b_{\delta b}(k)$ is given by the \citet{ShawRudd_2012} parameterisation, that is
\begin{equation}
b_{\delta b}(k)^2=\frac{1}{2}\left[\mathrm{e}^{-k / k_f}+\frac{1}{1+\left(g k / k_f\right)^2}\right],
\end{equation}
where $k_f = 9.4\,\mathrm{Mpc}^{-1}$ and $g = 0.5$ are constant with redshift and fitted to match high-resolution simulations \citep{Aubert2015_EMMA,GorceIlic_2020}.

\subsection{Power spectra of CMB imprints of reionisation}\label{subsec:signatures_ps}

With these two elements in hand, that is a parameterisation for both the reionisation history $x_e(z)$ and its morphology through $P_{ee}(k,z)$, one can derive the angular power spectra of the kSZ signal, the $\tau$ fluctuations, and the EoR-induced $B$-modes. 

\subsubsection{Thomson optical depth}

The Thomson optical depth represents the fraction of CMB photons absorbed by free electrons along the line-of-sight $\los$, and, as such, is directly related to the number density of free electrons in the IGM $n_e(\los, z)$ such that \citep{Planck2018}: 
\begin{equation}
\label{eq:def_tau}
\begin{aligned}
\tau(\los, z) &=\,\sigma_\mathrm{T} \int_{t(z)}^{t_0} \mathrm{d} t'\ n_e\left(\los, t'\right)\\
&= c\,\sigma_\mathrm{T} \int_{0}^{z} \frac{\mathrm{d} z'}{H(z')} \ \frac{x_e\left(\los,z'\right)}{(1+z')}\, n_b\left(\los, z'\right)\, ,
\end{aligned}
\end{equation} 
where $\sigma_T$ is the Thomson cross-section, $n_b(z)=n_{b,0}(1+z)^3$ is the average number density of baryons, $n_e = x_e n_b$, and $H$ is the Hubble parameter. 
The spatial fluctuations in the optical depth consequently write
\begin{equation}\label{eq:dtau}
\begin{aligned}
  \delta \tau(\los, z) &\equiv \frac{\tau(\los, z)}{\bar{\tau}(z)}-1 \\ 
  & = n_{b,0}\sigma_T \int \drm \eta' \bar{x}_e\left(\eta^{\prime}\right) \left(1+z^{\prime}\right)^2\delta_e\left(\los, \eta^{\prime}\right),
\end{aligned}
\end{equation}
where the integral is performed over the comoving distance $\eta$, $\bar{x}_e$ is spatially averaged and $\delta_e$ is the free electron overdensity, such that $P_{ee}(k,z)\propto\vert \tilde{\delta}_e(k,z)\vert^2$ -- the tilde denoting a spatial Fourier transform. Using the Limber approximation (valid for $\ell \gtrsim 10$) and in the limit of small fluctuations ($\delta \tau \ll \bar{\tau} \ll 1$), one can write the power spectrum of the $\tau$ spatial fluctuations as
\begin{equation}
\label{eq:cl_tau_Pee}
\begin{aligned}
C^{\tau\tau}_\ell &\simeq \bar{n}_{b,0}^2\sigma_\mathrm{T}^2 \int  \frac{\mathrm{d} \eta'}{\eta'^2}\, P_{ee}\left(k=\ell/\eta', z'\right) (1+z')^4\bar{x}_e(z')^2.
\end{aligned}
\end{equation}
Depending on the limits of the integral, one can isolate the contribution from patchy reionisation or from low-redshift large-scale structures, in a similar fashion to what is usually done for the kSZ angular power spectrum.

\subsubsection{The kSZ power spectrum}

The derivation of the kSZ angular power spectrum given the power spectrum of the free electron density fluctuations has been detailed in \cite{GorceIlic_2020}. Hence, here, we limit ourselves to the main derivations.

The CMB temperature anisotropies coming from the scattering of CMB photons off clouds of free electrons with a non-zero bulk velocity $\bm{v}$ relative to the CMB rest-frame along the line of sight $\los$ write
\begin{equation}\label{eq:dkSZ}
\begin{aligned}
\delta T_\mathrm{kSZ}(\los, z)&=\frac{\sigma_T}{c} \int_0^z \mathrm{d} \eta \,\frac{\mathrm{e}^{-\tau(\los, z)}}{(1+z)}  n_e(\los, z)\, \bm{v} \cdot \los\\
&= \frac{1}{c} \int_0^z\mathrm{d} \eta \, \frac{\mathrm{d}\tau}{\mathrm{d}\eta} \,\frac{\mathrm{e}^{-\tau(\los, z)}}{(1+z)}   \bm{v} \cdot \los\\
&=\sigma_T \int_0^z \frac{(1+z)^2}{H(z)} \frac{\mathrm{~d} z}{(1+z)} \mathrm{e}^{-\tau(\los, z)} n_e(\los, z)\, \bm{v} \cdot \los.
\end{aligned}
\end{equation}
Under the Limber approximation, the assumption that the velocity power spectrum is a biased linear matter power spectrum, and the omission of third and fourth order correlation terms \citep{GorceIlic_2020}, the kSZ angular power spectrum writes:
\begin{equation}
\begin{aligned}
\label{eq:def_C_ell_pkSZ}
C_\ell^\mathrm{kSZ} & \propto \int \frac{\bar{n}_e(z)^2}{(1+z)^2}\, \Delta_{B,e}^2(\ell/\eta,z)\, \exp^{-2 \tau(z)}\, \eta(z) \, \frac{\mathrm{d} \eta}{\mathrm{d} z}\, \mathrm{d} z ,\\
& = \frac{8 \pi^2}{(2\ell+1)^3} \frac{\sigma_T^2}{c^2} \int \frac{\bar{n}_e(z)^2}{(1+z)^2}\, \Delta_{B,e}^2(\ell/\eta,z)\, \exp^{-2 \tau(z)}\, \eta(z) \, \frac{\mathrm{d} \eta}{\mathrm{d} z}\, \mathrm{d} z.
\end{aligned}
\end{equation}
 We have $P_{B,e}$, the power spectrum of the curl component of the momentum field $\mathbf{q}_{B,e}$, such that $(2\pi)^3 P_{B,e}\, \delta_D(\kvec-\kvec')= \langle \qvec_{B,e}(\kvec)\ \qvec_{B,e}^*(\kvec')\rangle$ where $\delta_D$ is the Dirac delta function and the asterisk denotes a complex conjugate\footnote{We define the dimensionless cross-spectrum, for two fields $a$ and $b$ at a certain wave number $k$, as $\Delta^{2}_{\mathrm{a,b}}(k)\equiv k^{3} P_{\mathrm{a,b}}(k)/(2\pi^{2})$.}. We have
\begin{equation}
\label{eq:full_delta_B}
\begin{aligned}
  \frac{\langle \qvec_{B,e}(\kvec)\ \qvec_{B,e}^*(\kvec') \rangle}{(2\pi)^3{\delta_D}(|\kvec - \kvec'|)} \equiv & \frac{2\pi^2}{k^3} \Delta^2_{B,e}(k,z) \\
  = & \frac{1}{(2\pi)^3} \int \drm^3 k'\,  \left[ (1-\mu^2)\, \Pee (|\kvec-\kvec'|)\, P_{vv}(k') \right. \\ & \left. - \frac{(1-\mu^2)\, k'}{|\kvec-\kvec'|}P_{ev}(|\kvec-\kvec'|) \, P_{ev}(k') \, \right],
\end{aligned}
\end{equation}
where where $\mu = \kvecunit \cdot \kvecunit'$ and the $z$-dependencies have been omitted for readability. The free electrons density-velocity cross-spectrum $P_{ev}$ is taken to be
\begin{equation}
P_{v e}(k, z) =\frac{f \dot{a}(z)}{k} b_{\delta e}(k, z) P_{\delta \delta}^{\operatorname{lin}}(k, z)
\end{equation}
where $a$ is the scale factor, $f$ the linear growth rate and the bias is defined by the ratio $b_{\delta e}(k, z)^2 \equiv P_{e e}(k, z) / P_{\delta \delta}(k, z)$, computed from the analytical forms of both $P_{ee}$ and $P_{\delta \delta}$.

\subsubsection{EoR-induced $B$-modes}

\begin{figure}
    \centering
    \includegraphics[width=\columnwidth]{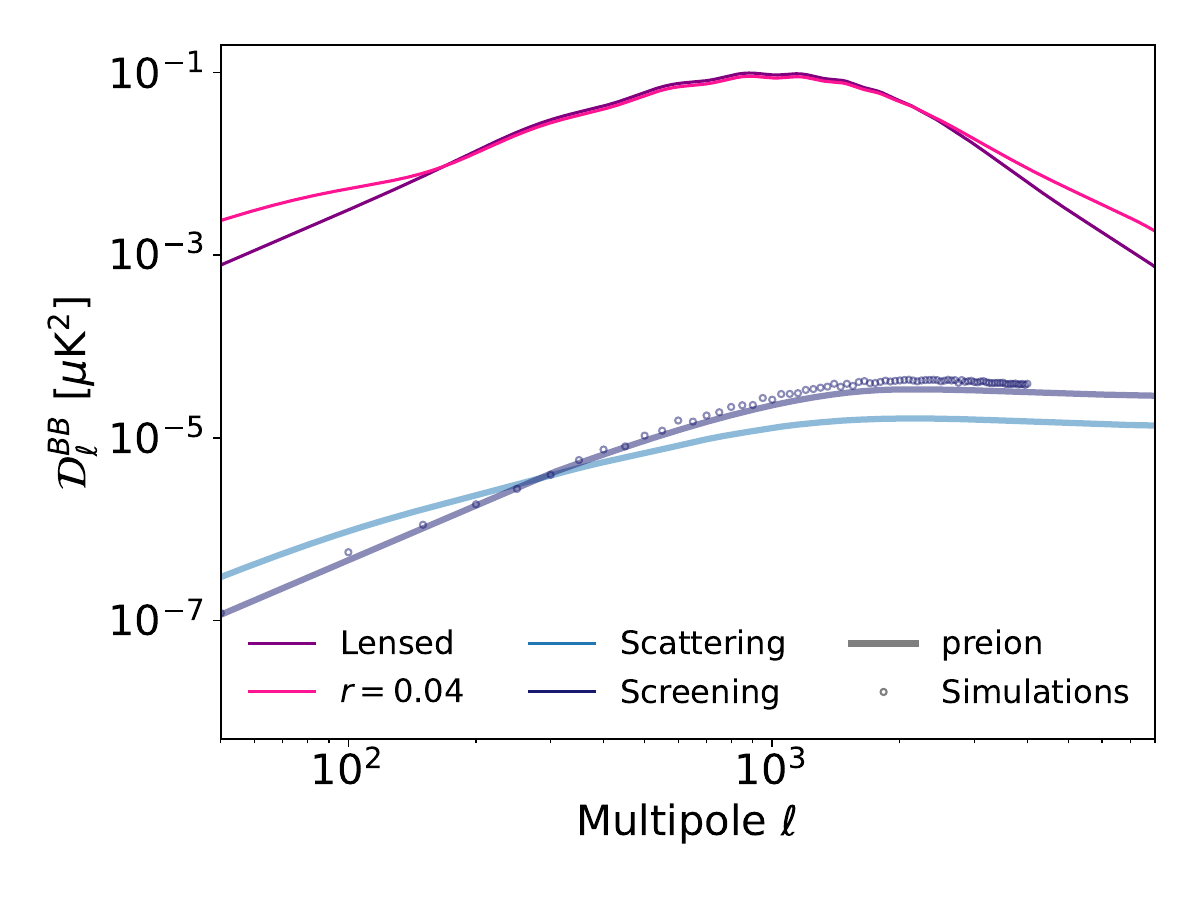}
    \caption{Components to the $BB$ angular power spectrum of CMB polarisation anisotropies. The primordial $B$-modes, with (in pink, $r=0.04$) and without (in purple), derived with \texttt{CAMB}, dominate. EoR-induced polarisation anisotropies shown in blue, from scattering and screening, have about 100~times lower amplitude. Results obtained given a $C_\ell^{\tau\tau}$ from the \texttt{preion} analytical computation and from simulations (see text), are compared.}
    \label{fig:bb_components}
\end{figure}

There are two mechanisms linked to reionisation and producing $B$-modes: screening and scattering of the quadrupole fluctuations during reionisation \citep{Hu_2000}, shown in Fig.~\ref{fig:bb_components} in dark and light blue, respectively. From the figure, it is clear that the anisotropies due to gravitational lensing dominate EoR-induced anisotropies: The angular power spectrum of the former is 5 to 50\,000 times larger than the latter, with the discrepancy being largest on the largest scales.

The screening on the primary anisotropies can be written as 
\begin{equation}
\label{eq:screening_maps}
\begin{aligned}
 T(\hat{\mathbf{n}})& =e^{-\tau(\hat{\mathbf{n}})} T^{(\mathrm{rec})}(\hat{\mathbf{n}}) \\
 (Q \pm i U)(\hat{\mathbf{n}}) &=e^{-\tau(\hat{\mathbf{n}})}(Q \pm i U)^{(\mathrm{rec})}(\hat{\mathbf{n}})
\end{aligned}
\end{equation}
where $T$ is the temperature fluctuation, and $Q$ and $U$ are the polarisation Stokes parameters. Primary anisotropies are inhomogeneously screened because of patchy reionisation and the resulting anisotropies in the optical depth\footnote{Since the gravitational lensing kernel is maximum at $z\sim 2$, that is long after the end of Hydrogen reionisation, the screening from the Epoch of Reionisation applies to primary anisotropies, stemming from recombination, denoted with a $\mathrm{(rec)}$ superscript. We ignore the concurring effect of lensing and screening, as we focus in the reminder of this work on scales where screening is dominated by the effect of patchy reionisation.}. 
The power spectrum of the induced $B$-mode anistropies is a convolution of $E$ anisotropies from recombination and $\tau$ power spectra \citep{DvorkinSmith_2009}:
\begin{equation}
\label{eq:screening_bb_full}
C_{\ell}^{B B(\mathrm{scr})}=e^{-2 \bar{\tau}} \int \frac{d^2 \ell^{\prime}}{(2 \pi)^2} C_{\ell^{\prime}}^{E E(\mathrm{rec})} C_{\left|\ell-\ell^{\prime}\right|}^{\tau \tau} \sin ^2\left(2 \phi_{\ell^{\prime}}\right).
\end{equation}
On large scales (above the acoustic scale, $\ell \lesssim 300$), because of the low amplitude of $C_\ell^{EE}$, this convolution can be approximated by a simple integral over both terms and the contribution is limited to white noise. On small scales (below the damping scale, $\ell \gtrsim 2000$), the screening power spectrum follows $C_\ell^{\tau\tau}$. In our model, these two asymptotic behaviours are connected through interpolation. On Fig.~\ref{fig:bb_components}, we compare the power obtained from applying Eqs.~\eqref{eq:screening_maps} for a given $C_\ell^{\tau\tau}$ to Gaussian random fields with the appropriate primary power spectra (dotted line), to what our analytical model predicts with the asymptotic behaviours of Eq.~\eqref{eq:screening_bb_full} (thick solid line). We find a good match between the two approaches, however, the analytical approximation underestimates the power on scales $\ell \gtrsim 1000$ by about 20-25\%.

Another contribution comes from fluctuations in the electron density generating new polarisation at the time of reionisation.
As can be seen on Fig.~\ref{fig:bb_components}, most of the large-scale power comes from these scattering $B$-modes ($\ell \lesssim 200$), whose angular power spectrum writes \citep{RoyLapi_2020}:
\begin{equation}
\label{eq:cl_bb_Pee}
\begin{aligned}
 C_{\ell}^{B B(\mathrm{sca})} 
 & =\frac{3 \sigma_{\mathrm{T}}^2 \bar{n}_{p, 0}^2}{100} \int \frac{\mathrm{d} \eta'}{\eta'^2}\, P_{ee}\left(\ell/\eta', z'\right) \exp^{-2\tau(\eta')}  \bar{x}_e(z')^2 (1+z')^4 Q_\mathrm{rms}^2,\\
 & \simeq \frac{3}{100} \exp^{-2\bar{\tau}}  Q_\mathrm{rms}^2 C_\ell^{\tau\tau}.
\end{aligned}
\end{equation}
Here, $Q_\mathrm{rms}$ is the root-mean-square of the primary temperature quadrupole. Although $Q_\mathrm{rms}$ should depend on redshift, since the power spectrum of the primordial fluctuations is almost scale-invariant, we take $Q_\mathrm{rms}=17~\mu\mathrm{K}$ constant, equal to its value during reionisation \citep{Hu_2000, DvorkinSmith_2009, RoyKulkarni_2021}. For the \citet{Planck2018} value of $\bar{\tau}$, the pre-factor of Eq.~\eqref{eq:cl_bb_Pee} is about $8~\mu\mathrm{K}^2$, such that the scattering $B$-modes power spectrum is a factor 8 larger than the Thomson optical depth power spectrum. Note that the amplitude of the $BB$ spectrum can vary by a factor two depending on the value of $Q_\mathrm{rms}$, which ranges between 15 and $22\,\mu\mathrm{K}$, depending on the cosmology. For this component, we also find in Fig.~\ref{fig:bb_components} a good match between simulations and the analytical approximations used to derive the angular power spectrum in Eq.~\eqref{eq:cl_bb_Pee}.\\

The formalism described in this Section allows us to derive the angular power spectra of EoR-induced anisotropies contributing to CMB observations, given a reionisation history (described by $z_\mathrm{re}$ and d$z$) and morphology (described by log$\alpha_0$ and $\kappa$).
 
\subsection{Detectability}

In this section, we describe the framework used to compare the EoR observables, obtained following Sec.~\ref{subsec:signatures_ps}, with the CMB primary signal, other secondary anisotropies (such as gravitational lensing), and to telescope sensitivities and sample variance.
We obtain the theoretical primary and lensed CMB temperature and polarisation power spectra with the Boltzmann integrator \texttt{CAMB}. We generate noise power spectra and cosmic variance following the specifications of various space- and ground-based experiments (Table~\ref{tab:forecasts}). In practice, there would be additional contributions from beam and calibrations errors, which we do not consider here\footnote{An option is to add $5\%$ beam uncertainty and $1\%$ temperature calibration uncertainty as done in \citet{zahn_2012_spt}.}.

\begin{table*}[]
    \centering
    \caption{Specifications of the different experiments considered in our forecasts. For all experiments, we average the noise and beam over the different frequency channels covered by the instrument where the primary CMB signal dominates other temperature anisotropies, that is, 90-145~GHz. Here, the map noise corresponds to temperature measurements. Note that, because we use the Limber approximation in our derivations, we do not compute signal below $\ell=10$.}
\begin{tabular}{l|ccccc|cc}
    \multirow{2}{3em}{Telescope} & \multicolumn{2}{c}{SO} & \multicolumn{2}{c}{CMB-S4} & \multirow{2}{*}{CMB-HD} & \multirow{2}{*}{LiteBIRD} & \multirow{2}{*}{PICO} \\
     & LAT & SAT & LAT & SAT &  &  &  \\
    \hline \hline
    $f_\mathrm{sky}$ &  0.4 & 0.1 & 0.4 & 0.03 & 0.5 & 0.5 & 0.5 \\
    $\Theta_\mathrm{FWHM}$ [arcmin] & 1.5 & 17.0 & 1.0 & 23.0 & 0.33 & 30 & 7.9 \\
    Map noise [$\mu$K-arcmin]  & 6.3 & 2.1 & 1.0 & 1.0 & 0.75 & 1.2 & 1.1\\
    $\ell_\mathrm{min}$ & 1000 & 30 & 30 & 20 &  1000 & 1 & 1 \\
    $\ell_\mathrm{max}$ & 8000 & 300 & 4000 & 330 & 25\,000 & 200 & 4000 \\
    $\Delta \ell$ & 500 & 50 & 500 & 50 & 500 & 50 & 10 \\
    References & \multicolumn{2}{c}{(1) (2)}  & \multicolumn{2}{c}{(3, 4, 5)} &  () & (5) & (6) \\
\end{tabular}
    \tablebib{(1) \citet{ZhuBhandarkar_2021}; (2) \citet{Ade_2019}; (3) \citet{CMB-S4} (4) \citet{cmbs4_white} (5) \citet{AbazajianAddison_2022}; (4) \citet{SehgalAiola_2020}; (5) \citet{LiteBIRD}; (6) \citet{PICO}.}
    \label{tab:forecasts}
\end{table*}

\subsubsection*{Cosmic and sample variance}

Given a power spectrum $\mathcal{D}_\ell$, including the contribution from primary signal and foregrounds, sample variance errors are given by
\begin{equation} \label{eq:cosmic_variance}
  \frac{\Delta \mathcal{D}_\ell}{\mathcal{D}^\mathrm{obs}_\ell} =\sqrt{\frac{2}{(2\ell+1)f_\mathrm{sky}}} 
\end{equation} 
where $\mathcal{D}^\mathrm{obs}$ corresponds to the total observed spectrum (including primary and secondary anisotropies, as well as noise, see below), and $f_\mathrm{sky}$ is the fraction of the sky observed by the instrument (Table~\ref{tab:forecasts}).

\subsubsection*{Noise bias}

The noise power spectrum for temperature fluctuations is given, for each $\ell$, by\footnote{This result is for a $C_\ell$. Our results are presented in terms of $\mathcal{D}_\ell \equiv \ell(\ell+1)C_\ell /2\pi$.}:
\begin{equation}
\label{eq:noise_ps}
    N_{\ell}^{TT} = \sigma_0^2 \ \mathrm{exp} \left[ \frac{\ell(\ell+1)\, \Theta_\mathrm{FWHM}^2}{8\ln 2} \right],
\end{equation}
where $\sigma_0$ is the map noise, given in $\mu$K-steradian and $\Theta_\mathrm{FWHM}$ is the beam full width half maximum, given in radian, for a beam considered Gaussian. 

\subsubsection*{Noise bias on the optical depth power spectrum}

Given the way patchy screening modifies the CMB polarisation and temperature anisotropies, as described in Eqs.~\eqref{eq:screening_maps}, we need to define a quadratic estimator to reconstruct the contribution of patchy screening to the observed power spectra. The proper noise to take into account when quantifying the detectability of $C_\ell^{\tau\tau}$ is the noise bias from this quadratic estimator.

Let $X$ and $Y$ be CMB maps -- either temperature $T$ or polarisation $E$ or $B$, that are screened by the same patchy optical depth $\tau(\hat{\mathbf{n}})$.
The screening modulates the amplitude of both maps: 
\begin{equation}
\langle X(\boldsymbol{l}_1)Y(\boldsymbol{l}_2)\rangle_{\mathrm{fixed} \, \tau} = f_{XY}(\boldsymbol{l}_1, \boldsymbol{l}_2) \tau(\boldsymbol{L})
\end{equation}
for $\boldsymbol{l}_1 \neq \boldsymbol{l}_2$ and  $\boldsymbol{L} = \boldsymbol{l}_1 + \boldsymbol{l}_2$. This $f_{XY}$ function describes the coupling between the different scales and depends on the fields considered (see \citet{SuYadav_2011} for an exhaustive list). For polarisation only, in the case of the $EB$ cross spectrum which gives the highest signal to noise for patchy reionisation \citep{DvorkinSmith_2009}\footnote{We confirm this result, first mentioned in \citet{DvorkinSmith_2009}: With our models, the signal-to-noise ratio is about 100 times larger for $EB$ than for any other map combination, across all telescopes considered.}, the response is given by
\begin{equation}
f_{EB}^\tau(\boldsymbol{l}_1, \boldsymbol{l}_2) =  \left(\bar{C}_{\ell_1}^{E E}-\bar{C}_{\ell_2}^{B B}\right) \sin 2\left(\phi_{\ell_1}-\phi_{\ell_2}\right),
\end{equation}
where the $\bar{C}_{l}^{EE}$ and $\bar{C}_l^{BB}$ are the primary (unlensed and unscreened) CMB polarisation power spectra. We can define a generic estimator of $\tau$ as
\begin{equation} \label{eq:tau_estimator}
\hat \tau_{XY}(\boldsymbol{L}) = \frac{1}{R_{XY}(\boldsymbol{L})} \int_{\boldsymbol{L}=\boldsymbol{l}_1 + \boldsymbol{l}_2} \bar{X}_{\boldsymbol{l}_1} \bar{Y}_{\boldsymbol{l}_2} W_{XY}(\boldsymbol{l}_1, \boldsymbol{l}_2) \;
\end{equation}
with $\bar{X}$ and $\bar{Y}$ the inverse-variance filtered CMB fields, $W_{XY}$ the weight of the estimator (also called the filter $F_{XY}^\tau$) and $R_{XY}(\boldsymbol{L})$ the normalisation which ensures the estimator is unbiased. In practice, in the $EB$ case, the optimal form of the filter is
\begin{equation}
W_{E B}^\tau\left(\boldsymbol{l}_1, \boldsymbol{l}_2\right) = F_{E B}^\tau\left(\boldsymbol{l}_1, \boldsymbol{l}_2\right)=\frac{f_{E B}^\tau\left(\boldsymbol{l}_1, \boldsymbol{l}_2\right)}{\left(\bar{C}_{\ell_1}^{EE} + N_{\ell_1}^{EE} \right)\left(\bar{C}_{\ell_2}^{BB} + N_{\ell_2}^{BB} \right)},
\end{equation}
in order to minimise the variance and to reduce the expectation value of the $\tau$ estimator to
\begin{equation}
\left\langle\hat{\tau}_{E B}\left(\vec{\ell}_1\right) \hat{\tau}_{E B}\left(\vec{\ell}_2\right)\right\rangle=(2 \pi)^2 \delta\left(\vec{\ell}_1, \vec{\ell}_2\right)\left[C_L^{\tau \tau}+\tilde{N}_{E B}^\tau(\vec{L})\right].
\end{equation}
We normalise our estimator by
\begin{equation}
R_{EB}(\boldsymbol{L}) = \int \frac{d^2 \vec{\ell}_1}{(2 \pi)^2} \frac{f_{E B}^\tau\left(\vec{\ell}_1, \vec{\ell}_2\right)^2}{\left(\bar{C}_{\ell_1}^{EE} + N_{\ell_1}^{EE} \right)\left(\bar{C}_{\ell_2}^{BB} + N_{\ell_2}^{BB} \right)},
\end{equation}
where $N_{\ell}^{XX}$ is the noise power spectrum in $X$ and the zeroth order bias for $\tau$ reconstruction is defined as $\tilde{N}^\tau_{EB}(\boldsymbol{L}) \equiv R_{EB}(\boldsymbol{L})^{-1}$. It is this noise we will use when quantifying the detectability of $C_\ell^{\tau\tau}$. 

The signal will be additionally distorted by gravitational lensing, inducing a non-negligible bias in the reconstructed screening signal \citep{SuYadav_2011}. The noise estimates performed as above do not take this effect into account and, therefore, correspond to a best-case scenario where the signal has been perfectly de-lensed. Various unbiased estimators have been proposed in the literature to simultaneously reconstruct $\tau$ fluctuations and the lensing potential $\phi(\los)$ with only marginal reductions in the SNR \citep{SuYadav_2011,FengHolder_2019, SchuttManiyar_2024}. In this paper, we limit our analysis to the unlensed case.\\

Taking both the cosmic variance and the noise components into account, we can compute the signal-to-noise ratio:
\begin{equation} \label{eq:snr_def}
\left(\frac{S}{N}\right)=\frac{\mathcal{D}_{\ell}}{\mathcal{D}_{\ell}^\mathrm{obs}}\sqrt{\frac{f_\mathrm{sky}}{2} (2 \ell+1)\Delta \ell},
\end{equation}
and the cumulative SNR, appropriate when looking to recover only the amplitude of the spectrum, is the sum of the above equation over all observed multipoles $\ell$ (Table~\ref{tab:forecasts}).

\subsection{Forecast}

To test the constraining potential of each CMB observable of reionisation, we use an Monte-Carlo Markov Chain (MCMC) sampler to estimate the probability of recovering our model parameters given our model and a mock observational dataset obtained with different telescopes.
Here, we use a simple Gaussian likelihood, and the prior we have on each model parameter is flat, with bounds listed in Table~\ref{tab:fid_params}. 

To generate our mock data, we use the fiducial parameter values listed in Table~\ref{tab:fid_params} and compute the CMB observables following the methodology described in the previous paragraphs.
We fix the cosmology to \citet{Planck2018} values. We add error bars to the measurements following Eqs.~\ref{eq:cosmic_variance} and \ref{eq:noise_ps}, although these expressions correspond to infinitely thin $\ell$-bins (with a Dirac-like window function). To approximate the fact that the power is measured on a bin and not at an exact multipole, we average the errors over each bin by dividing the amplitude of the error by $\sqrt{\Delta \ell}$ where $\Delta \ell$ is the width of the bin. 

\section{Results: Power spectra} \label{sec:results_ps}

\begin{figure*}
   \centering
   \includegraphics[width=\textwidth]
   {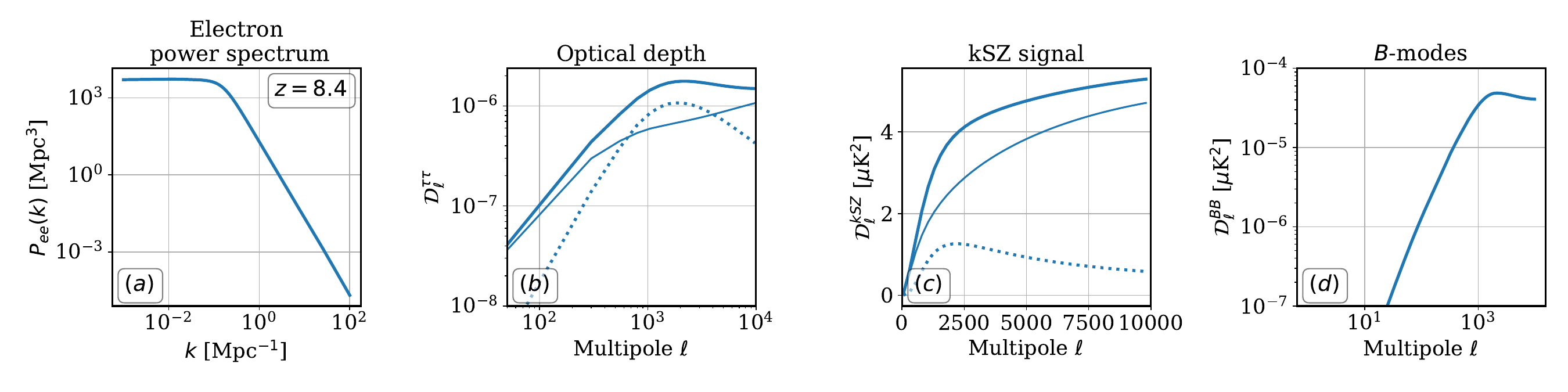}
    \caption{Propagation of our physical model to various reionisation observables. \textit{(a)} The electron power spectrum at $z=8.4$, the basis of the model (not an observable). \textit{(b)} Angular power spectrum of the Thomson optical depth fluctuations. \textit{(c)} Angular power spectrum of the kinetic Sunyaev-Zel'dovich effect. \textit{(d)} Angular power spectrum of the EoR-induced $B$ polarisation anisotropies (see also Fig.~\ref{fig:bb_components}). In \textit{(b)} and \textit{(c)}, the dotted line represents the contribution from reionisation redshifts, whilst the thin solid line represents the post-reionisation signal.}
         \label{Fig:fiducial_spectra}
   \end{figure*}

We present in Fig.~\ref{Fig:fiducial_spectra} the optical depth, kSZ, and EoR-induced $B$-modes derived from the electron power spectrum $P_{ee}(k,z)$ (leftmost panel) for the set of parameters given in Table~\ref{tab:fid_params}. We compare our results with previous works and investigate the dependency of our observables with the parameters of the model, that is, $z_\mathrm{re}$, $z_\mathrm{end}$, $\alpha_0$, and $\kappa$ (Fig.~\ref{Fig:param_dep}), over a range of plausible values\footnote{The minimal and maximal possible values of each parameter are taken from the results of \citet{GorceIlic_2020}, where the $P_{ee}$ parameterisation was fitted on a range of reionisation simulations.}.

\begin{table}[]
    \caption{Model parameter values used for our fiducial results (second row) and miminum and maximum values as limits of the prior range in the forecast and the latin hypercube sample (Sec.~\ref{sec:potential}).}
    \label{tab:fid_params}
    \centering
    \begin{tabular}{l|cccc}
         Parameter & $z_\mathrm{re}$ & d$z$ & log$\alpha_0$ & log$\kappa$ \\ \hline \hline
         Value & 7.0 & 1.5 & 3.7 & -1.0 \\ \hline
         Min. & 6.0 & 0.10 & 2.5 & -2.0\\
         Max. & 10.0 & 4.0 & 4.5 & -0.5\\
    \end{tabular}
    \tablefoot{An additional constraint d$z=z_\mathrm{re}-z_\mathrm{end}\geq 0.1$ is enforced in the sampling.}
\end{table}

\subsection{The optical depth power spectrum}\label{subsec:3_tau}

The panel $(b)$ of Fig.~\ref{Fig:fiducial_spectra} presents the angular power spectrum of the Thomson optical depth fluctuations, obtained following Eq.~\eqref{eq:cl_tau_Pee}. The $\mathcal{D}_\ell$ amplitude is of order $10^{-8}$ at $\ell=100$ and $10^{-6}$ at $\ell=1\,000$, which is on the lower hand of results in the literature \citep{DvorkinSmith_2009,FengHolder_2019,GuzmanMeyers_2021,RoyLapi_2020}, based on a simplified analytical model of reionisation\footnote{The difference stems from an amplitude difference in their $P_{\Delta x_e \Delta x_e} \equiv \bar{x}_e^2 P_{ee}$, which is 1 to 3 orders of magnitude larger than our values across the reionisation redshifts, whichever the model considered.}. On the other hand, our amplitude is similar what is obtained in \cite{SuYadav_2011}\footnote{The authors use simulations \citep{ZahnLidz_2007,ZahnMesinger_2011} based on the excursion set criterion \citep{FurlanettoZaldarriaga_2004b}, which is also at the basis of the 21CMFAST simulation package \citep{21cmFAST_2007, 21cmFAST_2011, 21cmfast_v3}.} and \citet{KramervanEngelen_2025} from semi-numerical simulation boxes, and in \citet{RoyKulkarni_2021} from hydrodynamic cosmological simulations post-processed for radiative transfer. Therefore, simple models of reionisation seem to overestimate the $\tau\tau$ power and could bias the detectability studies carried out in previous works \citep{RoyLapi_2018,BianchiniMillea_2022}.

\begin{figure*}
   \centering
   \includegraphics[width=\textwidth]
   {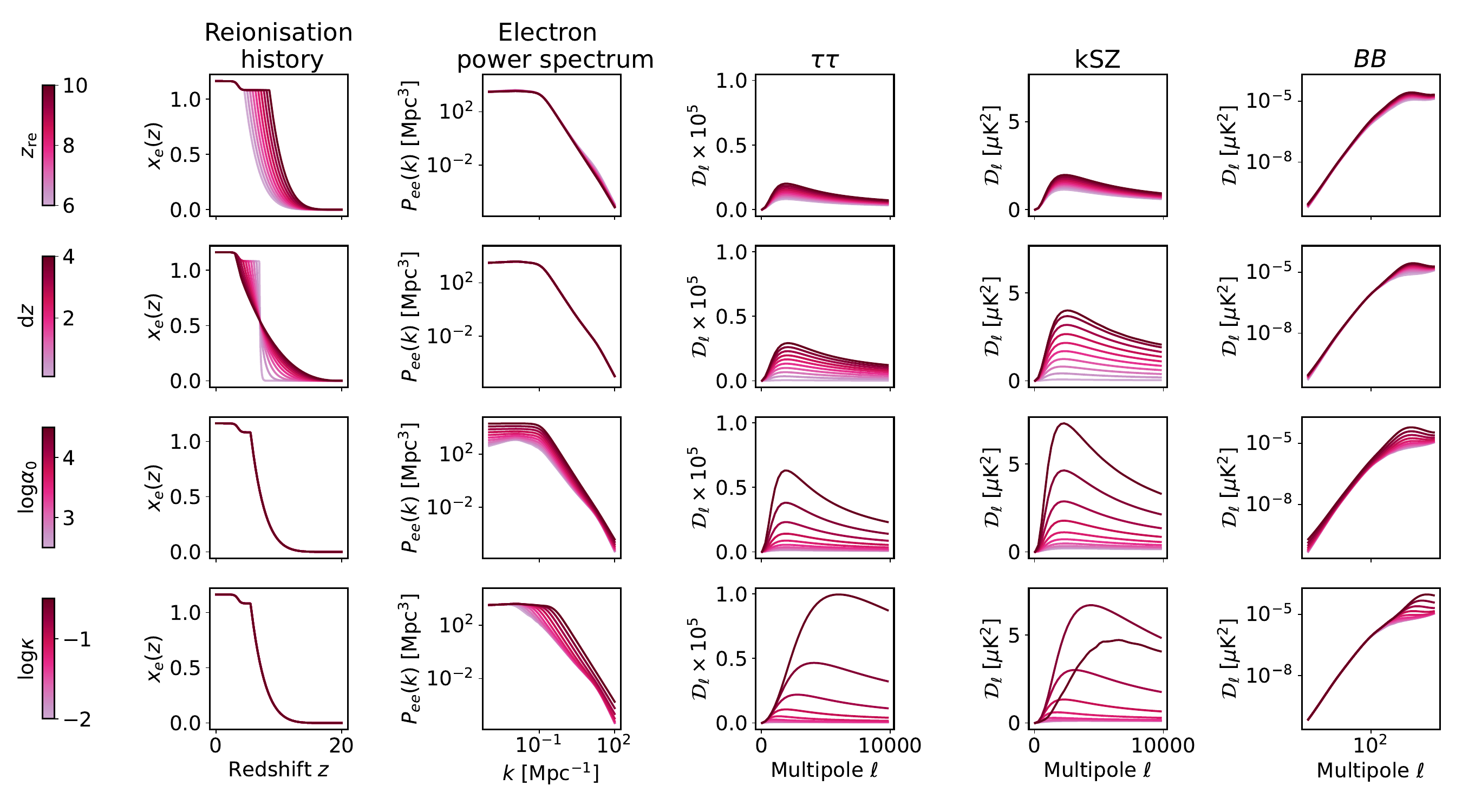}
    \caption{Dependence of the CMB observables of reionisation on the values of the model parameters ($z_\mathrm{re}$, $z_\mathrm{end}$, $\alpha_0$, $\kappa$), varied separately. The default values of the parameters correspond to the fiducial model of Fig.~\ref{Fig:fiducial_spectra}, with values listed in Table~\ref{tab:fid_params}.}
         \label{Fig:param_dep}
   \end{figure*}

We decompose the patchy $\tau$ signal into a low-redshift component, sourced by the large-scale structure of the Universe, and a high-redshift component, sourced mostly by the patchiness of reionisation, shown as a thin solid and a dotted line in Fig.~\ref{Fig:fiducial_spectra}b, respectively. We find the late-time $\tau\tau$ power to be flatter than its high-redshift counterpart -- similarly to kSZ. In contrast to the kSZ, the two components contribute in mostly equal parts to the total amplitude, with some variation depending on the multipole. The reionisation contribution dominates over $800 \lesssim \ell \lesssim 4000$, and is maximal (60\%) around $\ell=2000$, which corresponds to the multipole where the patchy kSZ signal is also maximal. Hence, the shape of the patchy $\tau\tau$ power spectrum traces the value of $\kappa$ (Fig.~\ref{Fig:param_dep}) and, in turn, the typical size of ionised bubbles during reionisation \citep[see also][]{GluscevicKamionkowski_2013}. 

As seen in Fig.~\ref{Fig:param_dep}, overall, the patchy\footnote{Note that we use `patchy' to qualify the reionisation contribution to both the kSZ and $\tau\tau$ signal, although there is also a small contribution due to matter fluctuations alone at these redshifts.} $\tau\tau $ power spectrum behaves in similar ways as the patchy kSZ with respect to model parameter values, that is, the amplitude of the signal is larger for longer and earlier reionisation scenarios, as well as for electron fields exhibiting more variance (larger $\alpha_0$) and smaller bubbles (larger $\kappa$). The dependence is of similar relative amplitude for all parameters: For a fractional increase of 10\% in log$\kappa$ (d$z$), both the kSZ and the $\tau\tau$ amplitudes at $\ell=2000$ increase by 61\% (9\%). Finally, we find their shapes to be almost identical, with the patchy kSZ spectrum peaking at slightly larger multipoles: For our fiducial model, it peaks around $\ell=2200$ whilst $\mathcal{D}_\ell^{\tau\tau}$ peaks at $\ell \sim 1800$, and has a stronger high-$\ell$ tail. These similarities mean that a measurement of the patchy kSZ spectrum could be used to place constraints on $C_\ell^{\tau\tau}$.

\subsection{The kSZ power spectrum}\label{subsec:3_ksz}

We generate the angular kSZ spectra with the emulator introduced in \cite{GorceDouspis_2022}, which reconstructs the spectra with errors smaller than $5\%$ (or $0.02~\mu\mathrm{K}^2$) at $\ell=3000$.
We limit this paragraph to a summary as the dependence of the patchy kSZ signal on the model parameters has already been detailed in \cite{GorceIlic_2020} \citep[see also][Fig.~A.2.]{GorceDouspis_2022}. Namely, we see in Fig.~\ref{Fig:param_dep} that the amplitude of the patchy kSZ power spectrum is mostly sensitive to the duration of reionisation (d$z$) and to its morphology, through $\alpha_0$ and $\kappa$. However, only $\kappa$, which can be assimilated to the smallest bubble size during reionisation, has an impact on the shape of the signal: Larger values of $\kappa$ (corresponding to smaller ionised bubbles) lead to the kSZ reaching its maximum at larger multipoles (or smaller angular scales). Hence, the kSZ `bump' can be used to constrain the typical size of ionised bubbles \citep[see also][]{zahn_2005,iliev_2007,mesinger_2012_kSZ}.

\subsection{The EoR-induced $B$-modes}\label{subsec:3_bb}

The amplitude of the $B$-polarisation anisotropies sourced by the EoR is about 3 to 4 orders of magnitude weaker than lensed $B$-modes but signal dominates over primordial gravitational waves on scales $\ell \gtrsim 400$ for $r= 10^{-3}$ (Fig.~\ref{fig:bb_components}). Across the parameter space, our models are 2.5 orders of magnitude lower than results obtained with the analytical model based on bubble size distributions \citep[BSD,][]{MortonsonHu_2007}. They are, however, in general agreement with the spectra presented in \citet{RoyLapi_2020}, based on high-dynamic-range radiative transfer simulations, with amplitudes at $\ell=200$ spanning the range $0.5-1 \times 10^{-5}~\mu\mathrm{K}^2$. We are not able to replicate the large signal obtained by the authors for a minimum halo mass of reionising galaxies of $M_\mathrm{min}=10^{11}M_\odot$ as this is an extreme model not covered by our parameter range. For the range of parameters considered, the maximum amplitude reached by a model is $2\times10^{-4}~\mu\mathrm{K}^2$, at $\ell=2000$. However, when jointly varying parameters, the signal goes as high as $10^{-3}~\mu\mathrm{K}^2$, but always on very small scales ($\ell\sim10^4$).

Overall, the dependence of $\mathcal{D}_\ell^{BB}$ on model parameters follows the one of $\mathcal{D}_\ell^{\tau\tau}$. Only $\alpha_0$ and, although more weakly, $z_\mathrm{re}$, have an impact on the large-scale component of the signal ($\ell < 100$), where scattering $B$-modes dominate. On smaller scales, and 
similarly to the other two observables, the amplitude of the signal increases for earlier and longer reionisation histories, as already observed in \citet{RoyLapi_2018}. 

We will confirm these qualitative dependencies on the model parameters with a parameter inference forecast in Sec.~\ref{subsec:4_forecast}.

\section{Results: Detectability \& Forecasts} \label{sec:potential}

In this section, we first look at the detectability of the three signals considered here with past, current, and future instruments (Sec.~\ref{subsec:4_detectability}). In Sec.~\ref{subsec:4_forecast}, we perform simple forecasts to assess the potential of each observable to constrain reionisation, first individually, and then jointly.

\subsection{Detectability} \label{subsec:4_detectability}

\begin{figure}
   \centering
   \includegraphics[width=.81\columnwidth]
   {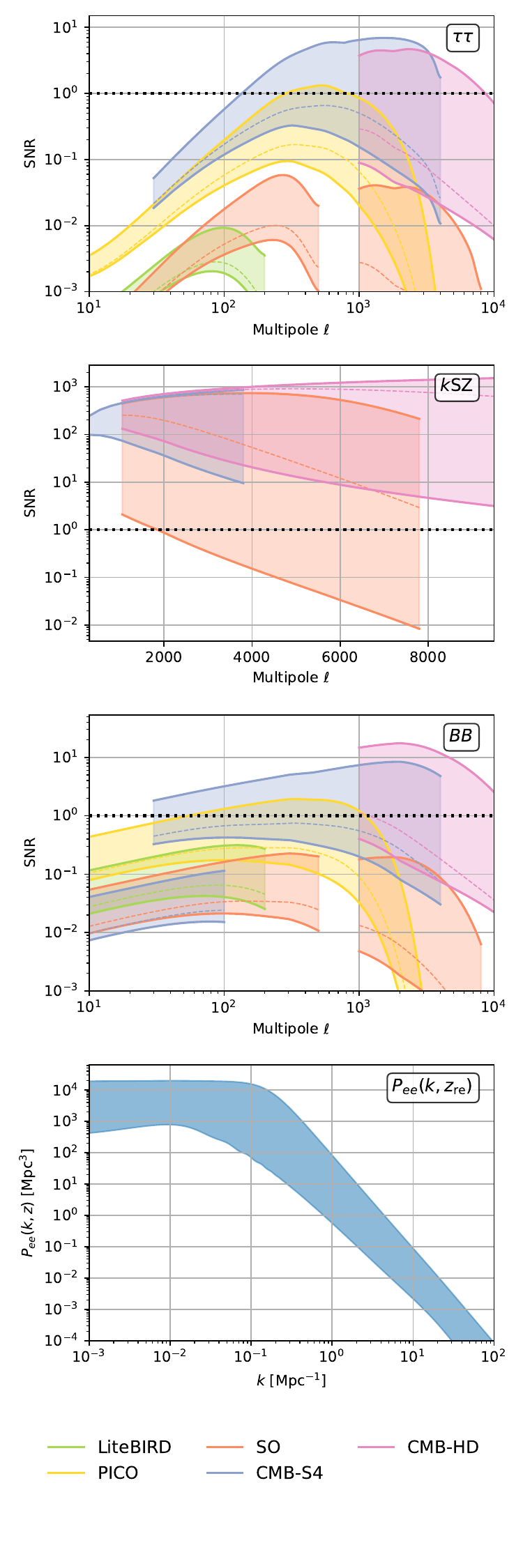}
    \caption{Range of signal-to-noise ratios of CMB observables of reionisation (considering bin width average), for 500 random values of the model parameters ($z_\mathrm{re}$, $z_\mathrm{end}$, $\alpha_0$, $\kappa$) taken from a latin hypercube (Table~\ref{tab:fid_params}) for different telescopes configurations (Table~\ref{tab:forecasts}). For the kSZ, only the patchy signal is considered, whilst the sum of all signal components is considered for $C_\ell^{\tau\tau}$ and $C_\ell^{BB}$. The range of models considered is represented in the lower panel, whilst our fiducial model is shown in dashed lines.}
         \label{Fig:snr_vs_models}
   \end{figure}

We compute the noise levels for a range of experiments according to Eq.~\eqref{eq:snr_def} and Table~\ref{tab:forecasts}. Because the SNR is model-dependent, we draw 500 parameter sets from a latin hypercube to homogeneously sample our 4-dimensional parameter space on a reasonable parameter range\footnotemark[5], given in Table~\ref{tab:fid_params}, and compare the dynamical range of our observables with telescope sensitivities. The result is a range of signal-to-noise ratios (SNR), shown in Fig.~\ref{Fig:snr_vs_models}. We list the resulting cumulative SNR for our fiducial model in Table~\ref{tab:snr}. All these results include the bin size $\Delta \ell$.

\begin{table}[h]
\caption{Cumulative signal-to-noise ratio (SNR) on the total (and EoR contribution to) $C_\ell^{\tau\tau}$ and $C_\ell^{BB}$ for various ground- and space-based CMB experiments (see Table~\ref{tab:forecasts} for specs).}
\label{tab:snr}
\centering
\begin{tabular}{lcccc}
\multirow{2}{6em}{\begin{center}
    Cumulative SNR
\end{center}} & \multicolumn{2}{c}{$C_\ell^{\tau\tau}$} & \multicolumn{2}{c}{$C_\ell^{BB}$} \\
\cmidrule(lr){2-3} \cmidrule(lr){4-5}
 & Total & Patchy & Total & Screen. \\
\midrule
LiteBIRD & -- & -- & 1.12 & 0.40 \\
PICO & 1.42 & 0.64 & 2.41 & 1.32 \\
SO-LAT & 0.01 & -- & 0.03 & 0.02 \\
SO-SAT & 0.01 & -- & 0.13 & 0.05 \\
CMB-S4-LAT & 2.16 & 1.10 & 2.72 & 1.72 \\
CMB-S4-SAT & -- & -- & 0.30 & 0.12 \\
CMB-HD & 1.14 & 0.58 & 4.28 & 2.97 \\
\bottomrule
\end{tabular}
\end{table}

For the patchy kSZ power spectrum, all experiments from Simons Observatory (SO) to CMB-HD are sensitive enough to measure the signal as a function of multipole on most of the multipole range, for all models: The SNR is larger than $10$ up to $\ell\sim4000$ for S4-LAT and CMB-HD. With SO-LAT observations, it will be possible to either measure or exclude most models, only the faintest being under the detection limit. These sensitivity estimates do not take into account the modelling of all anisotropies composing the observed CMB temperature power spectrum, including primary anisotropies which dominate the kSZ signal at $\ell \lesssim 1000$ and other foregrounds, such as the thermal SZ signal or simply the post-reionisation kSZ component, which dominate the patchy signal on all multipoles. Hence, the predicted SNR for, e.g., the SPT-3G camera of the South Pole Telescope would range between 5 and 10 across the multipole range, whilst the analysis of the latest data has not led to a detection, but to a $2\sigma$ upper limit \citep{ChaubalHuang_2026}.
Another option is to consider the kSZ spectra reconstructed from internal linear combination (ILC) of CMB temperature maps at different frequencies, as done in \citet{JainChoudhury_2023} following the cleaning method described in \citet{RaghunathanOmori_2023}. The resulting error bars are about $\sim 0.1~\mu\mathrm{K}^2$ for S4 \citep[Fig.~A1 of][]{RaghunathanOmori_2023}, which is about the same order of magnitude as our estimates, once accounting for the difference in bin width.
   
As described in Sec.~\ref{subsec:3_bb}, there are two regimes in the $BB$ spectrum: The signal is dominated by scattering from patchy reionisation at $\ell \lesssim 1300$, before screening $B$-modes take over at $\ell \gtrsim 2000$. 
For our fiducial model, only future experiments considered lead to a cumulative SNR above one, and up to 4.7 with CMB-HD (Table~\ref{tab:snr}). Our results show that measuring the signal on small scales is necessary to reach detection.
Amongst ground-based experiments, both CMB-S4-LAT and CMB-HD can measure the signal as a function of multipole for the brightest models, with CMB-HD only accessing the screening component of the signal. For the 50\% higher amplitude models, S4-LAT reaches a per-multipole SNR between 1 (at $\ell=30$) and 10 (at $\ell=2000$), accessing both the scattering and the screening components. The scattering $B$-modes could be measured from space: The cumulative SNR for PICO and our fiducial model is 2.63. Hence, a combination of ground- and space-based observations will lead to measurements of both the scattering and the screening $B$-modes. These numbers will slightly change depending on the value of $Q_\mathrm{rms}$, which controls the amplitude of scattering $B$-modes, dominant on the largest scales (Sec.~\ref{subsec:signatures_ps}). Across all the experiments considered here, the variation in cumulative SNR is $-10\%$ ($+25\%$) when $Q_\mathrm{rms}=15\,\mu\mathrm{K}$ ($22\,\mu\mathrm{K}$).
Since $C_\ell^{BB}$ is roughly proportional to $\mathrm{e}^{-2\bar{\tau}}C_\ell^{\tau\tau}$ across the multipole range, measuring the EoR-induced $B$-modes could be an interesting avenue to indirectly measure $C_\ell^{\tau\tau}$, or, reciprocally, to clean the observed $BB$ spectrum and access primordial gravitational waves. 

For optical depth fluctuations, all models and experiments lead to a SNR $<1$ on $\ell < 200$: The signal is too weak on very large scales to be detected and the LSS contribution to the optical depth fluctuations will be difficult to measure as a function of multipole with any of the telescopes considered here. A likely more promising approach to detecting the LSS contribution to $\tau$ fluctuations consists in targeting their cross-correlations with galaxy samples as proposed in \citet{SchuttManiyar_2024, HotinliHolder_2024} and done in \citet{CoultonSchutt_2024}. However, an amplitude detection of the auto-spectrum (LSS and patchy) would be possible with a telescope similar to the large aperture telescope (LAT) of CMB-Stage~4 (CMB-S4) as its cumulative SNR is about 2 (1.1 for the patchy contribution alone), but more difficult with PICO and CMB-HD whose cumulative SNR reaches 1.4 and 1.1, respectively, for the fiducial model. For brighter models, the per-multipole-bin SNR gets over 5 for CMB-S4-LAT and CMB-HD. In general, SNR are larger on smaller scales, where the signal is stronger. We note that the cumulative SNR for CMB-S4-LAT is larger than for CMB-HD (Table~\ref{tab:snr}) despite the better sensitivity of the latter (Table~\ref{tab:forecasts}). We find that this difference comes from the different multipole ranges covered by both experiments, and, more specifically, from the contribution of low multipoles ($30 \lesssim \ell \lesssim 1000$). Indeed, the QE presented in Sec.~\ref{subsec:3_tau} to reconstruct $\tau$ fluctuations relies on an integral over multipoles (Eq.~\ref{eq:tau_estimator}). 
Because of the lower amplitude of our $C_\ell^{\tau\tau}$, our forecast are more pessimistic than other results in the literature \citep{RoyLapi_2018, BianchiniMillea_2022}, especially on large scales. Additionally, \citet{JainMukherjee_2024} assume systematically larger $f_\mathrm{sky}$, leading to a factor about 2 difference in SNR. The results in Table~\ref{tab:snr} assume perfect delensing, which might not be achievable in practice: CMB-S4 is expected to reach 90\% delensing efficiency \citep{cmbs4_white}. To account for imperfect delensing, we add lensing residuals $C_\ell^{BB,\text{res.}} = (1-\epsilon) C_\ell^{BB,\text{lensing}}$ as an additional noise source in our quadratic estimator. We find for PICO, S4 and HD that, with $\epsilon=0.99$, the lensing residuals are below the instrumental noise, such that the reconstruction is only weakly impacted. For $\epsilon=0.90$, the cumulative SNR drops by 24\% for S4 and 6\% for PICO.

\begin{figure}
    \centering
    \includegraphics[width=.9\linewidth]{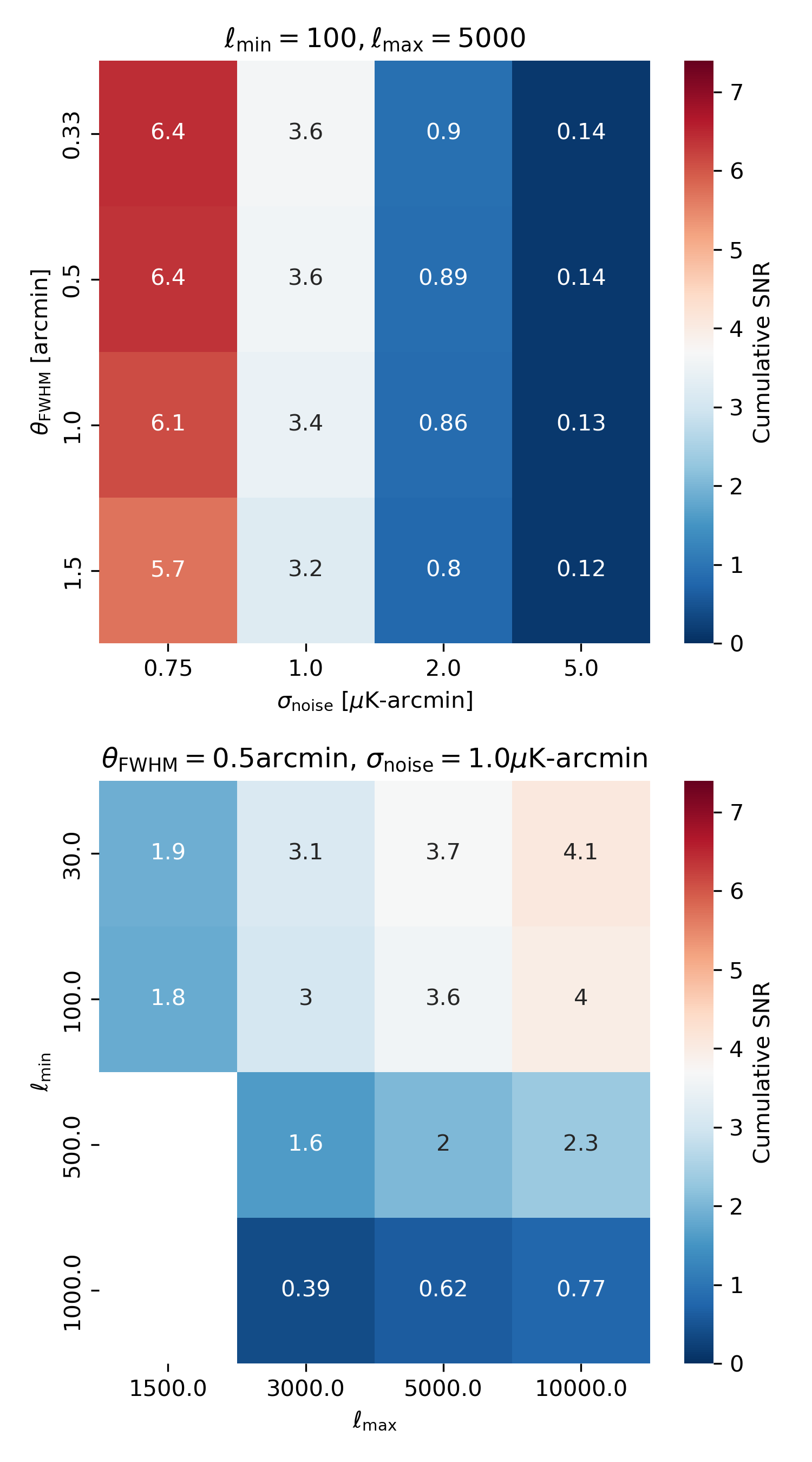}
    \caption{Cumulative SNR for $C_\ell^{\tau\tau}$ detection as a function of telescope specs, assuming full sky coverage ($f_\mathrm{sky}=1$) and our fiducial reionisation model (Table~\ref{tab:fid_params}).}
    \label{fig:csnr_vs_specs}
\end{figure}

We now investigate the telescope specifications which would be optimal to measure the $\tau\tau$ angular power spectrum. In Fig.~\ref{fig:csnr_vs_specs}, we present the cumulative SNR obtained when observing the whole sky ($f_\mathrm{sky}=1$) as a function of various observational and instrumental parameters: the minimum and maximum multiple accessible by the experiment, its map noise, and its beam full-width-half-maximum (FWHM). The reference case is taken as $100 \leq \ell \leq 5000$, $\theta_\mathrm{FWHM}=0.5'$ and a map noise of $1.0\,\mu\mathrm{K}-\mathrm{arcmin}$. We see that the beam size has little influence on the cumulative SNR, whilst a maximal map noise of $1\,\mu\mathrm{K}-\mathrm{arcmin}$ is required to reach a cumulative SNR above 3. Integrating over wider multipole ranges helps increase the cumulative SNR, reaching values above 3.5 as long as $\ell_\mathrm{min}\leq 100$ and $\ell_\mathrm{max}\geq 5000$. Therefore, as observed for CMB-HD, reaching higher multipoles does not help detection -- that is because the signal decreases significantly on small scales. 

\subsection{Constraining potential} \label{subsec:4_forecast}

\begin{figure}
    \centering
    \includegraphics[width=.8\linewidth]{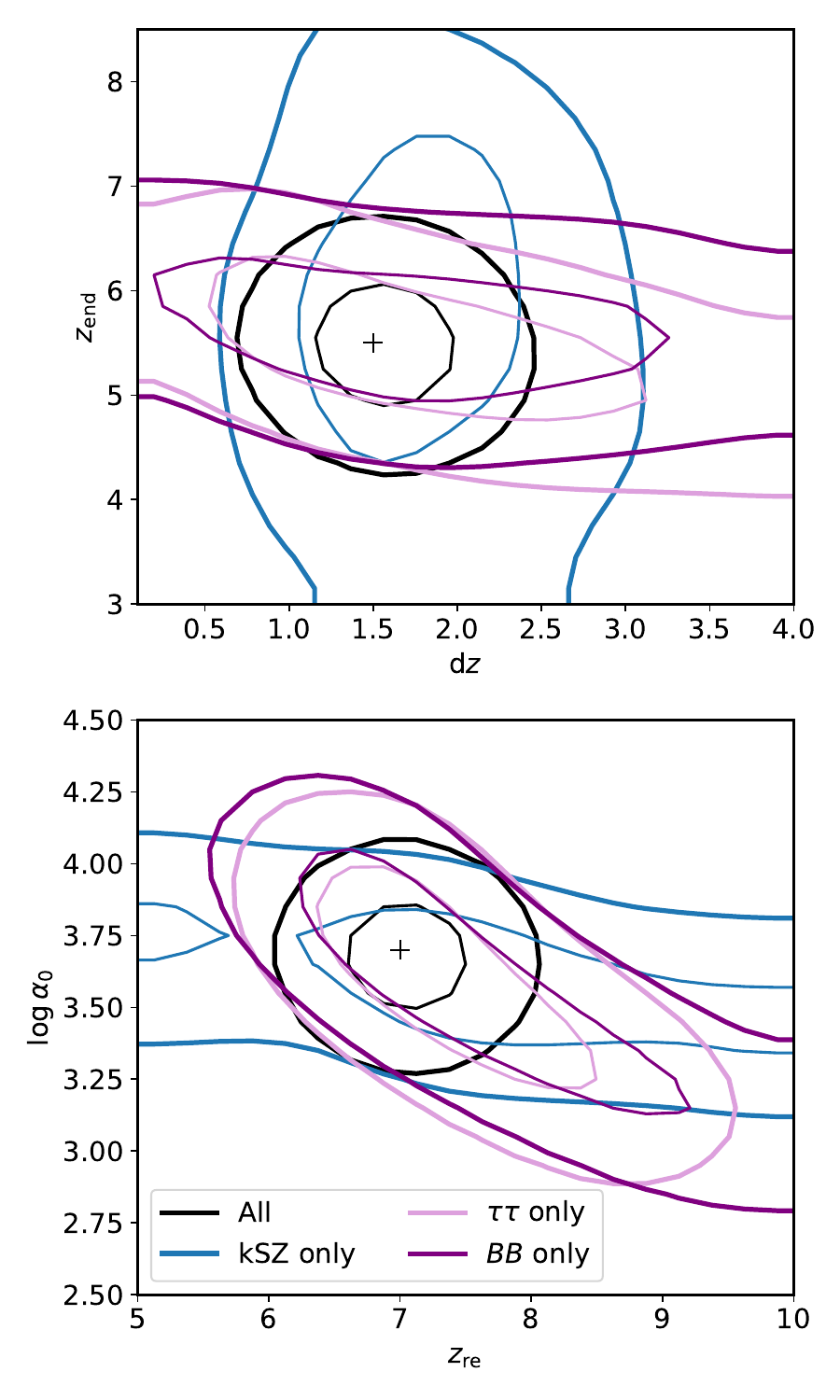}
    \caption{Posterior distributions of reionisation history and morphology parameters when fitting mock observations at cosmic variance limit of the patchy kSZ, the $\tau\tau$ power spectrum, the EoR-induced $B$-mode power spectrum, each individually, or a combination of the three (in black).}
    \label{fig:corner_cv_limited}
\end{figure}

In this section, we test the potential of each CMB observable independently, or a combination of the three, to constrain the four reionisation parameters of our model. We take, as our observables, measurements of the angular power spectra of the kSZ signal, the $\tau$ fluctuations, and the EoR-induced $B$-modes.
We first consider a scenario where the measurements of each observable are cosmic-variance limited. With this approach, we can test the intrinsic constraining power of each observable, independently of the sensitivity of instruments. In order to measure the intrinsic reionisation information embedded in each observable, we choose the following multipole ranges: $100 \leq \ell \leq 5000$ with $\Delta\ell=100$ for $C_\ell^{\tau\tau}$ and $C_\ell^{BB}$, $1000 \leq \ell \leq 8000$ with  $\Delta \ell = 500$ for the kSZ\footnote{Note that the mock data points fall exactly on the model: that is, we do not consider the contribution of random noise.}.
The results for chosen parameter combinations, shown as two-dimensional posterior probability distributions in Fig.~\ref{fig:corner_cv_limited}, confirm the qualitative findings of Sec.~\ref{sec:results_ps}. 
As expected, the kSZ signal can only provide tight constraints on the duration of reionisation (d$z$, upper panel vs. $z_\mathrm{re}$, lower panel), and $z_\mathrm{re}$ and $z_\mathrm{end}$ are strongly degenerated ($r^2=0.83$) when using kSZ data alone. This degeneracy does not exist in the other two observables, and both $C_\ell^{\tau\tau}$ and $C_\ell^{BB}$ give $\lesssim 10\%$ error constraints on $z_\mathrm{end}$. Reciprocally, $C_\ell^{\tau\tau}$ and $C_\ell^{BB}$ data are unable to constrain the reionisation duration d$z$ (upper panel), which is strongly degenerate with $z_\mathrm{re}$ ($r^2=0.82$ for both). Therefore, combining all observables breaks these degeneracies and leads to percent-level measurements of $z_\mathrm{re}$ and $z_\mathrm{end}$.
Regarding morphological parameters, all three observables can be used to measure the typical bubble size during reionisation, through the $\kappa$ parameter, with equivalent uncertainties ($\lesssim 3.5\%$). Combining the three observables provides the tightest constraints on $\log\alpha_0$, with a 1.5\% error bar, as $z_\mathrm{end}-\alpha_0$ have a different degeneracy direction in kSZ and $\tau\tau$. Overall, $C_\ell^{\tau\tau}$ and $C_\ell^{BB}$ have comparable constraining power, as already noted through their parameter dependencies in Sec.~\ref{subsec:3_bb} -- this is expected, given most of the constraining power comes from the small-scale screening $B$-modes, whose angular power spectrum follows $\mathrm{e}^{-2\bar{\tau}}C_\ell^{\tau\tau}$.
We note that combining the datasets can help mitigate incomplete measurements: For example, the lack of large-scale ($\ell \gtrsim 1000$) modes in a $C_\ell^{\tau\tau}$ measurement creates a degeneracy in $z_\mathrm{end}-\kappa$ which is broken by the addition of kSZ datapoints to the analysis. In conclusion, only a combination of the three observables gives an unbiased and narrow measurement of all four model parameters, as shown in black on the figure. 

In a second forecast, we find that upper limits on $C_\ell^{\tau\tau}$ can place upper limits on the reionisation duration d$z$ and morphological parameters as the brightest models are excluded. On the other hand, $z_\mathrm{re}$ remains fully unconstrained, as expected from Fig.~\ref{Fig:param_dep}. Such a measurement would be achievable with the LAT of CMB-S4, and equivalent constraints could be placed with an upper limit on $C_\ell^{BB}$. Indeed, as noted in Table~\ref{tab:snr}, the cumulative SNR in this set-up is 2.2, whilst it is 3.0 for $C_\ell^{BB}$.

\section{Conclusions}\label{sec:conclusions}

We have introduced a quick analytical model to evaluate the angular power spectra of the three CMB imprints of reionisation, from the patchy kSZ effect to fluctuations of the Thomson optical depth and associated EoR-induced $B$-modes (Sec.~\ref{sec:methods}), as a function of reionisation history ($z_\mathrm{end}$, $z_\mathrm{re}$) and morphology ($\alpha_0$, $\kappa$). We have compared the outputs of this model with commonly-used analytical models of reionisation and found that the latter tend to overestimate the patchy $\tau$ power by up to 100; in contrast, our power spectra are compatible with the outputs of both semi-numerical simulations and hydrodynamic cosmological simulations post-processed for radiative transfer (Sec.~\ref{sec:results_ps}), whilst requiring only a fraction of the simulations run time. 

We have shown that most of the reionisation information is encoded in the small scales of the CMB temperature and polarisation anisotropies power spectra: the signature bump of the patchy kSZ appears at $\ell \sim 2000$, $C_\ell^{\tau\tau}$ is dominated by the patchy EoR contribution at $1000 \lesssim \ell \lesssim 5000$ (Fig.~\ref{Fig:fiducial_spectra}), whilst most of the $B$-mode power on scales $\ell \gtrsim 500$ comes from patchy screening (Fig.~\ref{fig:bb_components}), whose angular power spectrum follows $\mathrm{e}^{-2\bar{\tau}}C_\ell^{\tau\tau}$.

We have further investigated the dependency of $C_\ell^{\tau\tau}$, $C_\ell^{BB}$, and $C_\ell^\mathrm{kSZ}$ on reionisation characteristics (morphology, history) and found that each tracer holds complementary information about reionisation. On one hand, in the cosmic-variance limit, the patchy kSZ is best at measuring the reionisation duration and bubble size but can only place lower limits on the reionisation endpoint. On the other hand, $C_\ell^{\tau\tau}$ and $C_\ell^{BB}$ have similar constraining power and can be used to measure the reionisation endpoint, but not duration. Overall, combining all three observables leads to the tightest measurements on all parameters by breaking degeneracies, such as between $\kappa$ and $z_\mathrm{end}$ (Fig.~\ref{fig:corner_cv_limited}). 

Considering cosmic variance, instrumental noise and QE reconstruction noise in the case of $C_\ell^{\tau\tau}$, we have looked into the detectability of each observable by current and future CMB space- and ground-based experiments (Table~\ref{tab:snr}): $C_\ell^{\tau\tau}$ ($C_\ell^{BB}$) amplitude can be detected on small scales by S4 (HD) with cumulative SNR 2 (5), and exclude extreme models (Fig.~\ref{Fig:snr_vs_models}). Scattering $B$-modes are better measured from space with a potential $2\sigma$ detection to be expected with PICO. We find that the minimal requirements to measure the patchy $\tau\tau$ spectrum with a SNR above 3 are $1~\mu\mathrm{K}-\mathrm{arcmin}$ noise and multipoles $100 < \ell < 5000$ being covered, whilst the beam size has little to no impact on the SNR (Fig.~\ref{fig:csnr_vs_specs}). It is interesting to note that, given $C_\ell^{BB} \propto \mathrm{e}^{-2\bar{\tau}}C_\ell^{\tau\tau}$, whether one signal or the other is first detected is equivalent in terms of reionisation constraints. Upper limits, achievable with CMB-S4, would be sufficient to exclude some models of reionisation, that is very long (d$z>3$) scenarios, or reionisation models based on very faint sources ($\kappa>0.3~\mathrm{Mpc}^{-1}$).

However, a measurement of either $C_\ell^{\tau\tau}$ or $C_\ell^{BB}$ would require precise de-lensing \citep[see, e.g.,][]{SuYadav_2011}, and be dependent on the presence of primordial gravitational waves (Fig.~\ref{fig:bb_components}), neither of which have been accounted for in this work. In Paper II, we will consider an alternative strategy to mitigate such contributions: Cross-correlations of these signals with another reionisation observable: the 21\,cm signal from cosmic hydrogen.

\begin{acknowledgements}
    The author thanks Louis Legrand and Giulio Fabbian for their help on lensing estimators, as well as Marian Douspis for useful discussions at the various stages of this project. The author also thanks the referee for their insightful review, which helped improve this manuscript.
    
    This work was supported by ANR PIA funding ANR-20-IDEES-0002. Additionally, the authors acknowledge support by Institut Pascal at Université Paris-Saclay during the Paris-Saclay Astroparticle Symposium 2024, with the support of the P2IO Laboratory of Excellence (program ``Investissements d’avenir" ANR-11-IDEX-0003-01 Paris-Saclay and ANR-10-LABX-0038), the P2I axis of the Graduate School of Physics of Université Paris-Saclay, as well as IJCLab, CEA, IAS, OSUPS, the IN2P3 master projet UCMN, and APPEC. The AstroParticle Symposium enabled fruitful discussions that were essential to the writing of this paper.
    
    This work made us of the $\texttt{plancklens}$ package\footnote{Available at \url{https://github.com/carronj/plancklens/}.} to derive quadratic estimators and $C_\ell^{\tau\tau}$ reconstructions \citep{Carron_2019, Planck2020_lensing}.
      
\end{acknowledgements}

\bibliographystyle{aa}
\bibliography{dtau_bib}

\end{document}